\documentclass[%
  reprint,
  showpacs,preprintnumbers,
  amsmath,amssymb,
  aps,
  prb,
]{revtex4-2}

\usepackage{graphicx}[]
\usepackage{dcolumn}
\usepackage{bm}
\usepackage{multirow}
\usepackage{braket}
\usepackage{color}
\usepackage{ulem}

\usepackage{url}
\usepackage{hyperref}

\definecolor{myblue}{RGB}{50,50,200}

\begin{document}

\preprint{APS/123-QED}

\title{
Widely-sweeping magnetic field--temperature phase diagrams for skyrmion-hosting centrosymmetric tetragonal magnets
}

\author{Satoru Hayami$^1$ and Yasuyuki Kato$^2$}
\affiliation{
$^1$Graduate School of Science, Hokkaido University, Sapporo 060-0810, Japan \\
$^2$Department of Applied Physics, the University of Tokyo, Tokyo 113-8656, Japan 
}
 
\begin{abstract}
We numerically investigate the stabilization mechanisms of skyrmion crystals under thermal fluctuations and external magnetic field in itinerant centrosymmetric tetragonal magnets. 
By adopting an efficient steepest descent method with a small computational cost, we systematically construct the magnetic field--temperature phase diagrams of the effective spin model derived from the itinerant electron model on a two-dimensional square lattice. 
As a result, we find that a square-type skyrmion crystal is stabilized by either or both of the high-harmonic wave-vector interaction and 
the biquadratic interaction under an external magnetic field.
Especially, we discover that the former high-harmonic wave-vector interaction can stabilize the skyrmion crystal only at finite temperatures when its magnitude is small. 
In addition to the skyrmion crystal, we also find other stable multiple-$Q$ states in the phase diagram.
Lastly, we discuss the correspondence of the phase diagrams between the effective spin model and the skyrmion-hosting material GdRu$_2$Si$_2$. 
The present results suggest a variety of multiple-$Q$ states could be driven by thermal fluctuations and external magnetic fields in centrosymmetric itinerant magnets. 
\end{abstract}
\maketitle

\section{Introduction}

Fermi surface instability in itinerant electron systems gives rise to abundant quantum states of matter in various fields of condensed matter physics, such as superconductivity and magnetism. 
For example, the nesting property of the Fermi surface has been identified as the origin of charge and/or spin density waves in metallic alloys, organic conductors, and other itinerant magnets~\cite{Gruner_RevModPhys.60.1129, Gruner_RevModPhys.66.1}. 
As such a Fermi surface nesting can ubiquitously occur for various lattice geometry depending on the electronic band structure, itinerant electron systems are an optimal platform to realize further exotic electronic ordered states. 

In particular, when the Fermi surfaces are nested by multiple different wave vectors, there is a chance of inducing the multiple-$Q$ states accompanied with noncollinear and noncoplanar spin configurations~\cite{hayami2021topological}. 
The most well-known examples are the triple-$Q$ noncoplanar (double-$Q$ coplanar) states found in the triangular and pyrochlore (checkerboard) lattice systems, where the perfect nesting of the Fermi surface occurs at a particular electron filling~\cite{Martin_PhysRevLett.101.156402, Chern_PhysRevLett.105.226403, Venderbos_PhysRevLett.109.166405}. 
Subsequently, similar multiple-$Q$ states have been revealed under various lattice structures when ($d-2$)-dimensional portions of the Fermi surfaces are connected by the multiple-$Q$ wave vectors in the extended Brillouin zone ($d$ is the spatial dimension): triangular~\cite{Akagi_JPSJ.79.083711, Kato_PhysRevLett.105.266405, Akagi_PhysRevLett.108.096401,Hayami_PhysRevB.90.060402,Hayami_PhysRevB.94.024424}, square~\cite{Agterberg_PhysRevB.62.13816,hayami_PhysRevB.91.075104,Hayami_PhysRevB.94.024424}, cubic~\cite{Hayami_PhysRevB.89.085124}, kagome~\cite{Barros_PhysRevB.90.245119, Ghosh_PhysRevB.93.024401}, honeycomb~\cite{Jiang_PhysRevLett.114.216402, Venderbos_PhysRevB.93.115108}, and Shastry-Sutherland lattices~\cite{Shahzad_PhysRevB.96.224402}.  
More recently, the concept of the multiple-$Q$ states induced by the Fermi surface instability has been extended to long-period magnetic structures, such as the double-$Q$ stripe state~\cite{Ozawa_doi:10.7566/JPSJ.85.103703,batista2016frustration} and the triple-$Q$ skyrmion crystal (SkX)~\cite{Ozawa_PhysRevLett.118.147205, Hayami_PhysRevB.99.094420, Eto_PhysRevB.104.104425, Eto_PhysRevLett.129.017201,kobayashi2022skyrmion}.  

The emergence of the multiple-$Q$ states in itinerant electron systems is intuitively understood from the competition between the negative bilinear exchange interaction and the positive biquadratic interaction in momentum space: The former originates from the Ruderman-Kittel-Kasuya-Yosida (RKKY) interaction to induce the single-$Q$ spiral instability~\cite{Ruderman, Kasuya, Yosida1957} and the latter originates from the higher-order RKKY interaction to lead to the multiple-$Q$ instability~\cite{Akagi_PhysRevLett.108.096401, Hayami_PhysRevB.95.224424}, both of which are characterized by effective long-range spin interactions in real space. 
Especially, an effective spin model incorporating the effect of the above interactions describes a similar multiple-$Q$ instability to that in the original itinerant electron model. 
As the computational cost for the effective spin model is much cheaper compared to that for the itinerant electron model, it can be used to investigate the multiple-$Q$ instabilities in various situations with different lattice structures and magnetic anisotropy. 
In fact, it was clarified that manifold multiple-$Q$ states appear as the ground state by analyzing the effective spin model, such as the SkX in the hexagonal~\cite{hayami2020multiple,Hayami_PhysRevB.103.054422,Hayami_PhysRevB.105.014408,Hayami_PhysRevB.105.184426,Hayami_PhysRevB.105.224411,Hayami_PhysRevB.105.224423}, tetragonal~\cite{Hayami_PhysRevLett.121.137202,hayami2018multiple,Su_PhysRevResearch.2.013160,Hayami_PhysRevB.103.024439}, trigonal~\cite{yambe2021skyrmion,hayami2022skyrmion}, and orthorhombic~\cite{Hayami_doi:10.7566/JPSJ.91.093701} lattice systems, the hedgehog crystal in the cubic lattice system~\cite{Okumura_PhysRevB.101.144416,Shimizu_PhysRevB.103.054427,hayami2021field,Kato_PhysRevB.104.224405,Shimizu_PhysRevB.103.184421,Okumura_doi:10.7566/JPSJ.91.093702}, and the meron--antimeron crystal in the hexagonal lattice system~\cite{Hayami_PhysRevB.104.094425}. 
These systematic investigations might be useful to understand the origin of the SkXs in the hexagonal compounds Gd$_2$PdSi$_3$~\cite{Saha_PhysRevB.60.12162,kurumaji2019skyrmion,sampathkumaran2019report,Kumar_PhysRevB.101.144440,paddison2022magnetic,Bouaziz_PhysRevLett.128.157206} and Gd$_3$Ru$_4$Al$_{12}$~\cite{chandragiri2016magnetic,Nakamura_PhysRevB.98.054410,hirschberger2019skyrmion,Hirschberger_10.1088/1367-2630/abdef9} and the tetragonal compounds GdRu$_2$Si$_2$~\cite{khanh2020nanometric,Yasui2020imaging,khanh2022zoology} and EuAl$_4$~\cite{Shang_PhysRevB.103.L020405,kaneko2021charge,takagi2022square,Zhu_PhysRevB.105.014423}, the hedgehog crystal in the cubic compounds MnSi$_{1-x}$Ge$_{x}$~\cite{tanigaki2015real,kanazawa2017noncentrosymmetric,fujishiro2019topological,Kanazawa_PhysRevLett.125.137202,Kanazawa_doi:10.7566/JPSJ.91.101002} and SrFeO$_3$~\cite{Mostovoy_PhysRevLett.94.137205,Ishiwata_PhysRevB.84.054427,Ishiwata_PhysRevB.101.134406,Rogge_PhysRevMaterials.3.084404,Onose_PhysRevMaterials.4.114420,yambe2020double}, the vortex crystal in the hexagonal compound Y$_3$Co$_8$Sn$_4$~\cite{takagi2018multiple}, and the bubble state in the tetragonal compound CeAuSb$_2$~\cite{Marcus_PhysRevLett.120.097201,Park_PhysRevB.98.024426,Seo_PhysRevX.10.011035,seo2021spin}.

On the other hand, the studies on the multiple-$Q$ instability against thermal fluctuations have still been limited compared to those in the ground state. 
As it was demonstrated that thermal fluctuations tend to enhance the stability of the multiple-$Q$ states in the localized spin model~\cite{Muhlbauer_2009skyrmion, Okubo_PhysRevLett.108.017206, Buhrandt_PhysRevB.88.195137, Hayami_PhysRevB.93.184413, Laliena_PhysRevB.96.134420, Laliena_PhysRevB.98.224407} and they induce the finite-temperature phase transitions between the multiple-$Q$ states in the itinerant electron model~\cite{Chern_PhysRevLett.109.156801, Barros_PhysRevB.88.235101, Hayami_10.1088/1367-2630/ac3683, hayami2021phase}, the appearance of further intriguing multiple-$Q$ states is also expected based on the effective spin model~\cite{Kato_PhysRevB.105.174413}. 
In addition, it is important to construct the magnetic phase diagram against not only the magnetic field at low temperatures but also at higher temperatures in order to further clarify the validity of the effective spin model for real materials. 

In the present study, we examine the magnetic field--temperature phase diagram of the effective spin model consisting of the momentum-resolved interactions with a particular emphasis on the stabilization of the square-lattice SkX in centrosymmetric itinerant magnets. 
To this end, we perform numerical calculations based on the steepest descent method, which enables us to efficiently find the optimal spin configurations in the thermodynamic limit~\cite{Kato_PhysRevB.105.174413}. 
We focus on two mechanisms of the square-lattice SkX: One is the positive biquadratic interaction~\cite{Hayami_PhysRevB.103.024439} and the other is the high-harmonic wave-vector interaction~\cite{Hayami_doi:10.7566/JPSJ.89.103702, Hayami_PhysRevB.105.174437}. 
By carrying out the calculations in a wide range of the two interaction parameters, we find their similarity and difference in their magnetic field and temperature dependence. 
The mechanism based on the high-harmonic wave-vector interaction tends to favor both the single-$Q$ and double-$Q$ states depending on the field, while that based on the biquadratic interaction tends to favor the double-$Q$ states irrespective of the magnetic field. 
Furthermore, we show that the former mechanism can induce the SkX only at finite temperatures by tuning the interaction. 
We also discuss the relevance to the skyrmion-hosting material GdRu$_2$Si$_2$. 
The results of our systematic investigation will be a reference to understanding the microscopic mechanism of the SkX-hosting tetragonal magnets in the wide range of the temperatures, such as GdRu$_2$Si$_2$~\cite{khanh2020nanometric,Yasui2020imaging,khanh2022zoology}, EuAl$_4$~\cite{Shang_PhysRevB.103.L020405,kaneko2021charge,takagi2022square,Zhu_PhysRevB.105.014423}. EuGa$_4$~\cite{zhang2022giant,Zhu_PhysRevB.105.014423}, EuGa$_2$Al$_2$~\cite{moya2021incommensurate}, Mn$_{2-x}$Zn$_x$Sb~\cite{Nabi_PhysRevB.104.174419}, and MnPtGa~\cite{ibarra2022noncollinear}.

The rest of this paper is organized as follows. 
In Sec.~\ref{sec: Model and method}, we introduce the effective spin model of the itinerant electron model on a square lattice, which has anisotropic bilinear and biquadratic interactions in momentum space. 
We describe two mechanisms to stabilize the SkX: the high-harmonic wave-vector interaction and the biquadratic interaction. 
We also outline the numerical method based on the steepest descent method~\cite{Kato_PhysRevB.105.174413}. 
Then, we present the magnetic field--temperature phase diagram while changing the high-harmonic wave-vector interaction and the biquadratic interaction under the out-of-plane field in Sec.~\ref{sec: Phase diagram under out-of-plane field} and the in-plane field in Sec.~\ref{sec: Phase diagram under in-plane field}. 
Finally, we compare the phase diagrams in the effective spin model with that in GdRu$_2$Si$_2$ in Sec.~\ref{sec: Comparison with skyrmion-hosting materials}. 
We conclude this paper in Sec.~\ref{sec: Summary}.

\section{Model and method}
\label{sec: Model and method}

\subsection{Model}

\begin{figure}[t!]
\begin{center}
\includegraphics[width=1.0 \hsize ]{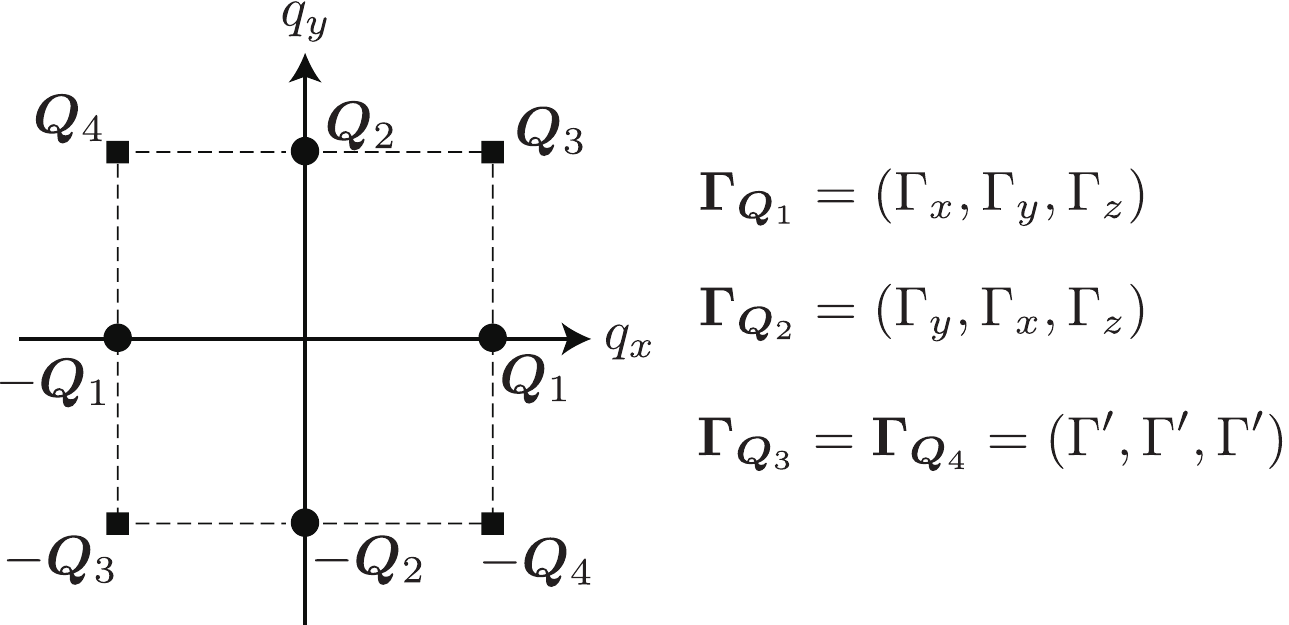} 
\caption{
\label{fig: model}
Momentum-resolved interactions $\bm{\Gamma}_{\bm{Q}_\nu}$ at $\bm{Q}_1=(Q,0)$, $\bm{Q}_2=(0,Q)$, $\bm{Q}_3=(Q,Q)$, and $\bm{Q}_4=(-Q,Q)$ with $Q=\pi/3$. 
}
\end{center}
\end{figure}

The square SkX in centrosymmetric magnets can emerge when considering the multi-spin interaction~\cite{Christensen_PhysRevX.8.041022, Hayami_PhysRevB.103.024439} and the high-harmonic wave-vector interaction~\cite{Hayami_doi:10.7566/JPSJ.89.103702, hayami2022multiple, Hayami_PhysRevB.105.104428, Hayami_PhysRevB.105.174437} in the effective spin model that originates from the itinerant electron model
or considering the bond-dependent anisotropy~\cite{Wang_PhysRevB.103.104408}, dipolar interaction~\cite{Utesov_PhysRevB.103.064414}, and the staggered DM interaction~\cite{hayami2022square} as well as the frustrated exchange interaction in the localized spin model.  
Among them, we focus on the stabilization of the square SkX in the former situation based on the effective spin model. 

Specifically, we consider the effective spin model on a two-dimensional square lattice under the point group $D_{4\rm h}$ in the following: 
\begin{align}
\label{eq: Ham}
\mathcal{H}=  &-J \sum_{\nu,\alpha,\beta}\Gamma^{\alpha\beta}_{\bm{Q}_\nu}S^{\alpha}_{\bm{Q}_\nu}S^{\beta}_{-\bm{Q}_\nu} \nonumber \\
&+\frac{K}{N} \sum_{\nu}\left(\sum_{\alpha,\beta}\Gamma^{\alpha\beta}_{\bm{Q}_\nu}S^{\alpha}_{\bm{Q}_\nu}S^{\beta}_{-\bm{Q}_\nu}\right)^2  
-  \sum_{j} \bm{H} \cdot  \bm{S}_{j}
\end{align}
where $S^\alpha_{\bm{Q}_\nu}$ is characterized by the wave vector $\pm\bm{Q}_1, \pm\bm{Q}_2, \cdots, \pm\bm{Q}_{N_{\bm{Q}}}$ and the spin component $\alpha,\beta=x,y,z$, which corresponds to the Fourier transformation of the classical localized spin $\bm{S}_{j}$ with $|\bm{S}_{j}|=1$:
\begin{align}
\bm{S}_{\bm Q} = \frac{1}{\sqrt{N}} \sum_j \bm{S}_j e^{-i {\bm Q}\cdot{\bm r}_j},
\end{align}
where $N$ represents the total number of sites and $\bm{r}_j=(r^x_j, r^y_j)$ denotes the position vector at site $j$.
We set the lattice constant as unity, and $r^x_j$ and $r^y_j$ are integers. 
The first and second terms in Eq.~(\ref{eq: Ham}) represent the bilinear and biquadratic spin interactions in momentum space, respectively; $\Gamma^{\alpha\beta}_{\bm{Q}_\nu}$ represents the anisotropic form factor depending on the wave vector $\bm{Q}_\nu$ and spin component $\alpha,\beta$. 
The third term represents the Zeeman coupling under an external magnetic field $\bm{H}$; we consider the $z$-directional field $\bm{H}=(0,0,H)$ in Sec.~\ref{sec: Phase diagram under out-of-plane field} and the $x$-directional field $\bm{H}=(H,0,0)$ in Sec.~\ref{sec: Phase diagram under in-plane field}. 

The effective spin model in Eq.~(\ref{eq: Ham}) is derived from the perturbation theory for the Kondo lattice model in the weak-coupling regime~\cite{Hayami_PhysRevB.95.224424,yambe2022effective}. 
The coupling constants in the first and second terms, $J>0$ and $K>0$, correspond to the lowest- and second-lowest-order contributions in terms of the Kondo coupling $J_{\rm K}$, respectively; for example, $J$ ($K$) is proportional to the second (fourth) order of 
$J_{\rm K}$.
We set $J=1$ as the energy unit of the model and treat $K$ as a phenomenological parameter. 
It is noted that we neglect the other four-spin interactions, e.g., $(S^{\alpha}_{\bm{Q}_\nu}S^{\beta}_{-\bm{Q}_\nu})(S^{\alpha'}_{\bm{Q}_{\nu'}}S^{\beta'}_{-\bm{Q}_{\nu'}})$ for $\nu \neq \nu'$, for simplicity~\cite{Akagi_PhysRevLett.108.096401, Ozawa_doi:10.7566/JPSJ.85.103703, Hayami_PhysRevB.95.224424}. 
The anisotropic form factor $\Gamma^{\alpha\beta}_{\bm{Q}_\nu}$ is determined by the spin--orbit coupling and the lattice symmetry in addition to the electronic band structure. 

For the dominant interaction channel at $\bm{Q}_\nu$, we consider the interactions at fourfold-symmetric wave vectors $\{\pm \bm{Q}_1=\pm(Q,0), \pm \bm{Q}_2=\pm(0,Q)\}$ with $Q=\pi/3$ by supposing that the nesting at $\bm{Q}_1$ and $\bm{Q}_2$ is important; $\bm{Q}_1$ and $\bm{Q}_2$ are related to the fourfold rotational symmetry of the tetragonal lattice structure. 
Then, $\Gamma^{\alpha\beta}_{\bm{Q}_\nu}$ is given as follows: $\bm{\Gamma}_{\bm{Q}_1}\equiv \Gamma^{\alpha\alpha}_{\bm{Q}_1}=(\Gamma_x, \Gamma_y, \Gamma_z)$ and $\bm{\Gamma}_{\bm{Q}_2}=(\Gamma_y, \Gamma_x, \Gamma_z)$, where $\Gamma^{\alpha\beta}_{\bm{Q}_\nu}=0$ for $\alpha \neq \beta$. 
We set $\Gamma_x=0.855$, $\Gamma_y=0.95$, and $\Gamma_z=1$ unless otherwise stated~\cite{khanh2022zoology}; $\Gamma_z> \Gamma_x, \Gamma_y$ means the easy-axis anisotropic interaction, which tends to favor the SkX and $\Gamma_y> \Gamma_x$ means the bond-dependent anisotropic interaction, which fixes the spiral plane onto the $yz$ ($xz$) plane for $\bm{Q}_1$ ($\bm{Q}_2$). 

Moreover, we consider the contribution from the high-harmonic wave vectors, i.e., $\pm \bm{Q}_3=\pm (\bm{Q}_1+\bm{Q}_2)$ and $\pm\bm{Q}_4=\pm(-\bm{Q}_1+\bm{Q}_2)$, since it can lower the energy to form the multiple-$Q$ state compared to the single-$Q$ state within the RKKY level~\cite{hayami2022multiple}. 
As we suppose that the interactions at $\bm{Q}_3$ and $\bm{Q}_4$ are smaller than those at $\bm{Q}_1$ and $\bm{Q}_2$, we set the isotropic form factor for simplicity; $\bm{\Gamma}_{\bm{Q}_3}=\bm{\Gamma}_{\bm{Q}_4}=(\Gamma', \Gamma', \Gamma')$. 
In the end, we investigate the instability toward the SkX while changing $K$ and $\Gamma'$. 
The wave vectors $\bm{Q}_1$--$\bm{Q}_4$ and their interactions are presented in Fig.~\ref{fig: model}.

\subsection{Method}

We investigate the optimal spin configurations of the effective spin model in Eq.~(\ref{eq: Ham}) at finite temperatures based on the steepest descent method with a set of self-consistent equations, which has been recently formulated~\cite{Kato_PhysRevB.105.174413}. 
In general, the effect of thermal fluctuations, i.e., the entropic effect, leads to the shrinking of the localized spin moment. 
As such fluctuations are brought about by the spatial correlation between the spins with a distance by the magnetic period in the classical spin model, 
we define an averaged spin for each sublattice in an $L \times L$ periodic array of the magnetic unit cell consisting of $\Lambda \times \Lambda$-site square cluster~\cite{comment_muc} 
 as 
\begin{align}
\bar{\bm{S}}_{\eta}= \frac{1}{L^2}\sum_{l} \bm{S}_{l,\eta}, 
\end{align}
where the site index ($j$) is redefined by a pair of numbers ($l,\eta$); $l$ and $\eta$ denote the indices of the magnetic unit cell and the sublattice site within the magnetic cell, respectively. 
With this setup, the linear dimension of the entire system is $L\Lambda$, and the total number of sites is $N=(L\Lambda)^2$;
the position vector $\bm{r}_\eta =(r^x_\eta, r^y_\eta)$ of sublattice $\eta$ is restricted within a magnetic unit cell: $r^x_\eta$, $r^y_\eta \in [0,\Lambda-1]$. 
In this paper, we consider the case of $Q=\pi/3$ that corresponds to $\Lambda=6$. 

Then, the partition function is calculated by~\cite{Kato_PhysRevB.105.174413} 
\begin{align}
Z= \int \left[ \Pi_{\eta} d\bar{\bm{S}}_{\eta} \rho (\bar{\bm{S}}_{\eta}) \right] e^{-\mathcal{H}/T}, 
\end{align}
where $\int d\bar{\bm{S}}_{\eta}$ means an integral over the unit ball ($|\bar{\bm{S}}_{\eta}| \leq 1$), $\rho (\bar{\bm{S}}_{\eta})$ is the density of states for $\bar{\bm{S}}_{\eta}$, and $T$ is the temperature (the Boltzmann constant is set to be unity). 
By taking the thermodynamic limit of $L \to \infty$ and using the steepest descent method, the resultant partition function is given by~\cite{Kato_PhysRevB.105.174413} 
\begin{align}
Z &\sim e^{L^2 G(\{\overline{\bar{S}^\alpha_{\eta}}\})}, \\
G(\{\bar{S}^\alpha_{\eta}\}) & = -\frac{\beta}{L^2} \mathcal{H} + \sum_\eta  V_\eta
\end{align}
with
\begin{align}
\label{eq: V}
V_\eta =
\ln \left[ \frac{4\pi \sinh v_0 (|\bar{\bm{S}}_{\eta}|)}{v_0 (|\bar{\bm{S}}_{\eta}|)}\right] - v_0 (|\bar{\bm{S}}_{\eta}|)|\bar{\bm{S}}_{\eta}| 
\end{align}
where $\{\overline{\bar{S}^\alpha_{\eta}}\}$ represents the saddle point that gives the maximum of $G(\{\bar{S}^\alpha_{\eta}\})$ and directly corresponds to the expectation value of each spin in the thermodynamic limit.
In Eq.~(\ref{eq: V}), $v_0(|\bar{\bm{S}}|)$ is determined by 
\begin{align}
\coth v_0(|\bar{\bm{S}}|) - \frac{1}{v_0(|\bar{\bm{S}}|)}=|\bar{\bm{S}}|. 
\end{align}
Once the saddle-point solution is obtained, the free energy is calculated via $-T \ln Z$. 
When several stable solutions are obtained for different initial spin configurations, we adopt the state with the lowest free energy. 

\section{Phase diagram under out-of-plane field}
\label{sec: Phase diagram under out-of-plane field}

\begin{figure*}[t!]
\begin{center}
\includegraphics[width=1.0 \hsize ]{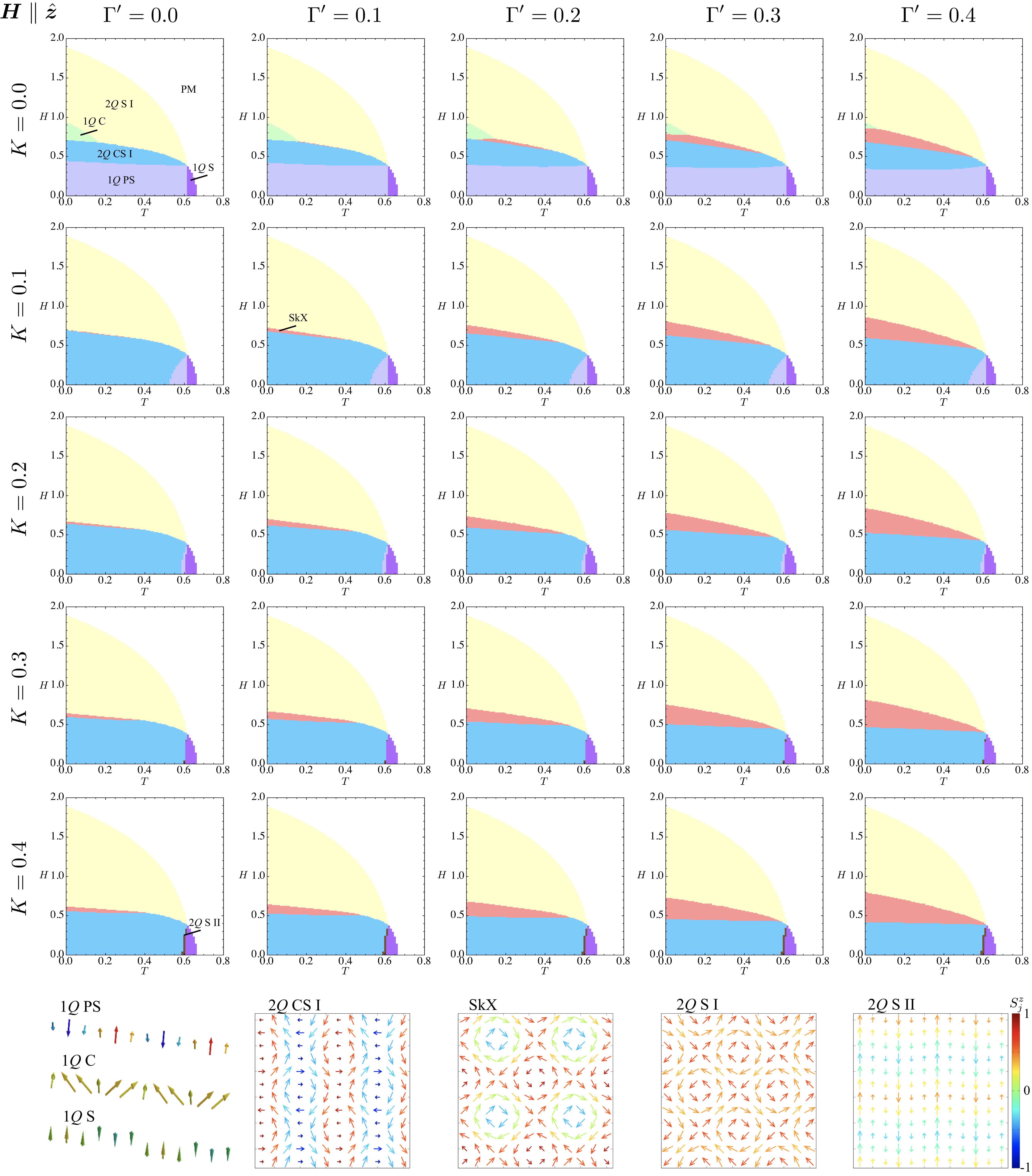} 
\caption{
\label{fig: PD_Hz}
Magnetic field($H$)--temperature ($T$) phase diagrams of the model in Eq.~(\ref{eq: Ham}) for $\bm{H} \parallel \hat{\bm{z}}$ with changing the biquadratic interaction $K$ and the high-harmonic wave-vector interaction $\Gamma'$.  
The schematic spin configurations appearing in the phase diagrams are presented in the bottom panel; the arrows represent the direction of the spin moments and their color shows the $z$-spin component. 
$1Q$ and $2Q$ represent the single-$Q$ and double-$Q$ states, respectively. 
PS, C, S, CS, and SkX mean proper-screw, conical, sinusoidal, chiral stripe, and skyrmion crystal, respectively. 
}
\end{center}
\end{figure*}

\break
In this section, we discuss the case when applying the magnetic field along the $z$ direction, i.e., $\bm{H}=(0,0,H)$. 
Figure~\ref{fig: PD_Hz} shows a collection of the magnetic field--temperature phase diagrams of the model in Eq.~(\ref{eq: Ham}) with changing $K$ by $\Delta K =0.1$ and $\Gamma'$ by $\Delta \Gamma' = 0.1$. 
In the wide range of parameters in terms of $K$ and $\Gamma'$, we obtain seven magnetic phases in addition to the paramagnetic (PM) state at high temperatures. 
We present the real-space spin configurations in each phase in the bottom panel of Fig.~\ref{fig: PD_Hz}, where the arrows represent the direction of the spin moments and their color shows the $z$-spin component.
We also list nonzero scalar chirality $\chi^{\rm sc}$ and nonzero $\bm{Q}_\nu$ components of the magnetic moments $m^{\alpha}_{\bm{Q}_\nu}$ in each phase in Table~\ref{table: OP_Hz}, which are given by 
\begin{align}
\chi^{\rm sc}&= \frac{1}{2 \Lambda^2} 
\sum_{\eta}
\sum_{\delta,\delta'= \pm1}
\delta \delta'
 \bar{\bm{S}}_{\eta} \cdot (\bar{\bm{S}}_{\eta+\delta\hat{x}} \times \bar{\bm{S}}_{\eta+\delta'\hat{y}}), \\
m^{\alpha}_{\bm{Q}_\nu}&= \frac{1}{\Lambda^2} \sqrt{\sum_{\eta,\eta'} \bar{S}^{\alpha}_{\eta}\bar{S}^{\alpha}_{\eta'}e^{i \bm{Q}_\nu \cdot (\bm{r}_\eta-\bm{r}_{\eta'})}},
\end{align}
where $\hat{x}$ ($\hat{y}$) represents a shift by lattice constant in the $x$ ($y$) direction.

\begin{table}[htb!]
\centering
\caption{
Scalar chirality $\chi^{\rm sc}$ and momentum-resolved magnetic moments $\bm{m}_{\bm{Q}_{\nu}}$ for $\nu=$1--4, in each magnetic phase for $\bm{H} \parallel \hat{\bm{z}}$. 
The subscript for $Q$ represents the index for the ordering vector. 
In the 2$Q$ S II phase, the spin configuration with $m^{x}_{\bm{Q}_{2}}$ and $m^{z}_{\bm{Q}_{1}}$ also gives the same energy. 
\label{table: OP_Hz}}
\renewcommand{\arraystretch}{2}
\begin{tabular}{lccccccccccc}\hline \hline
Phase  &$\chi^{\rm sc}$ & $m^{x}_{\bm{Q}_{1,2}}$ & $m^{y}_{\bm{Q}_{1,2}}$ & $m^{z}_{\bm{Q}_{1,2}}$ & $m^{x}_{\bm{Q}_{3,4}}$  & $m^{y}_{\bm{Q}_{3,4}}$ 
 & $m^{z}_{\bm{Q}_{3,4}}$  \\ \hline
1$Q$ PS & -- & -- & $1Q$ & $1Q$ & -- & -- & -- \\ 
1$Q$ S & -- & -- & -- & $1Q$ & --  & -- & --\\ 
1$Q$ C & -- & $1Q$ & $1Q$ & -- & --  & -- & --\\ 
2$Q$ CS & -- & $1Q_2$ & $1Q_1$ & $1Q$ & $2Q$  & -- & --  \\ 
2$Q$ S I & -- & $1Q_2$ & $1Q_1$ & -- & --  & -- & -- \\ 
2$Q$ S II & -- & -- & $1Q_1$ & $1Q_2$ & --  & -- & --\\ 
SkX & $\checkmark$ & $1Q_2$ & $1Q_1$ & $2Q$ & $2Q$ & $2Q$ & $2Q$   \\ 
\hline \hline
\end{tabular}
\end{table}

When $K=0$ and $\Gamma'=0$, the SkX does not appear in the phase diagram, as shown in the upper-left panel of Fig.~\ref{fig: PD_Hz}. 
Meanwhile, there are several double-$Q$ states in addition to the single-$Q$ states. 
By looking at the low-temperature region, the single-$Q$ proper-screw spiral (1$Q$ PS) state appears for low $H$, whose spiral plane lies on the plane perpendicular to $\bm{Q}_\nu$. 
For example, this state has nonzero components of $m^y_{\bm{Q}_1}$ and $m^z_{\bm{Q}_1}$ or $m^x_{\bm{Q}_2}$ and $m^z_{\bm{Q}_2}$. 
With increasing $H$, the 1$Q$ PS state continuously turns into the double-$Q$ chiral stripe I (2$Q$ CS I) state with the additional sinusoidal modulation at $\bm{Q}_2$ ($\bm{Q}_1$) for the spiral state with $\bm{Q}_1$ ($\bm{Q}_2$).
Thus, this state is characterized by nonzero $m^y_{\bm{Q}_1}$, $m^z_{\bm{Q}_1}$, and $m^x_{\bm{Q}_2}$ or $m^x_{\bm{Q}_2}$, $m^z_{\bm{Q}_2}$, and $m^y_{\bm{Q}_1}$. 
In addition, this 2$Q$ CS I state has small but nonzero amplitudes of $m^x_{\bm{Q}_3}$ and $m^x_{\bm{Q}_4}$ due to a superposition of the spin density waves at $\bm{Q}_1$ and $\bm{Q}_2$. 
Reflecting a noncoplanar spin texture, this state accompanies the density wave in terms of the scalar chirality, although its uniform component becomes zero. 
The 2$Q$ CS I state changes into the single-$Q$ conical (1$Q$ C) state with a jump of $m^\alpha_{\bm{Q}_\nu}$, whose spiral plane lies on the $xy$ plane, i.e., $m^x_{\bm{Q}_1} \neq m^y_{\bm{Q}_1} \neq 0$ or $m^x_{\bm{Q}_2} \neq m^y_{\bm{Q}_2} \neq 0$. 
With further increasing $H$, the 1$Q$ C state is replaced by the double-$Q$ sinusoidal I (2$Q$ S I) state with a jump of $m^\alpha_{\bm{Q}_\nu}$, whose spin configuration consists of the two sinusoidal waves of the $y$-spin ($x$-spin) component along the $\bm{Q}_1$ ($\bm{Q}_2$) direction with the same amplitude; $m^y_{\bm{Q}_1} = m^x_{\bm{Q}_2}$. 
Meanwhile, there is no $\bm{Q}_\nu$ component in the $z$ spin. 
The 2$Q$ S I state continuously turns into the fully-polarized state denoted by the PM in the phase diagram in Fig.~\ref{fig: PD_Hz}. 

When considering the effect of finite temperatures, two characteristic points are found. 
One is that the 1$Q$ C state is rapidly destabilized compared to the other three states, 1$Q$ PS, 2$Q$ CS I, and 2$Q$ S I states.  
Especially, one finds that the region of the 1$Q$ C state is replaced by that of the 2$Q$ S I state with increasing $T$, which implies that the entropic effect tends to favor the sinusoidal superposition rather than the single spiral. 
The other is the appearance of the single-$Q$ sinusoidal (1$Q$ S) state in the low-field and high-temperature region. 
This is attributed to the easy-axis exchange interaction in the model, i.e., $\Gamma_z> \Gamma_x, \Gamma_y$. 
It is noted that these two points have been also found in the frustrated spin model with the dipolar interaction~\cite{Utesov_PhysRevB.103.064414}, which suggests that these are feasible features irrespective of the short-range and long-range interactions. 

The appearance of various magnetic phases in the phase diagram is due to the presence of the easy-axis and bond-dependent anisotropic exchange interactions at $\bm{Q}_1$ and $\bm{Q}_2$. 
Indeed, only the 1$Q$ C state appears in the phase diagram when setting $\Gamma_x=\Gamma_y=\Gamma_z$. 
However, only the anisotropic exchange interactions are not enough to stabilize the SkX, at least, in the present parameters; $\Gamma_x=0.855$, $\Gamma_y=0.95$, and $\Gamma_z=1$. 
In the following, we show that the SkX emerges by additionally taking into account $\Gamma'$ and $K$ in Secs.~\ref{sec: Case of high-harmonic wave-vector interaction_Hz} and \ref{sec: Case of biquadratic interaction_Hz}, respectively. 
We also discuss the magnetic field--temperature phase diagram under both $\Gamma'$ and $K$ in Sec.~\ref{sec: Case of both interactions_Hz}. 

\subsection{Case of high-harmonic wave-vector interaction}
\label{sec: Case of high-harmonic wave-vector interaction_Hz}

The phase diagrams at $K=0$ for $\Gamma'=0.1$--$0.5$ are shown in the top panel of Fig.~\ref{fig: PD_Hz}. 
When introducing $\Gamma'$, the SkX appears in the vicinity region among the $2Q$ CS I, $1Q$ C, and $2Q$ S I, whose stability region extends with increasing $\Gamma'$.  
The SkX is characterized by a double-$Q$ superposition of two spiral waves along the $\bm{Q}_1$ and $\bm{Q}_2$ directions, as shown in the bottom panel of Fig.~\ref{fig: PD_Hz}. 
Although the in-plane spin configuration is similar to that in the $2Q$ S I, the SkX exhibits the additional $z$-spin modulation, which results in a nonzero uniform scalar chirality $\chi^{\rm sc}$ causing the topological Hall effect. 
In addition, the SkX has the intensities in both in-plane and $z$ spin components at high-harmonic wave vectors $\bm{Q}_3$ and $\bm{Q}_4$. 

Remarkably, the instability of the SkX is found at finite temperatures rather than zero temperature. 
Especially, the SkX only appears at finite temperatures for small $\Gamma'=0.1$ and $\Gamma'=0.2$. 
This indicates that thermal fluctuations tend to favor the SkX in the effective spin model with the momentum-resolved interactions~\cite{Kato_PhysRevB.105.174413}, similar to that in the frustrated spin model with the real-space competing interactions~\cite{Okubo_PhysRevLett.108.017206,Mitsumoto_PhysRevB.104.184432, Mitsumoto_PhysRevB.105.094427}. 
On the other hand, the present SkX phase is replaced by the other phases before entering the paramagnetic phase when increasing $T$, which is in contrast to the frustrated spin model, where the SkX region touches the paramagnetic region.  
This result suggests that the entropy in the SkX phase is larger than that of the 2$Q$ CS I phase, while it is smaller than that of the 2$Q$ S I state. 

\begin{figure*}[t!]
\begin{center}
\includegraphics[width=1.0 \hsize ]{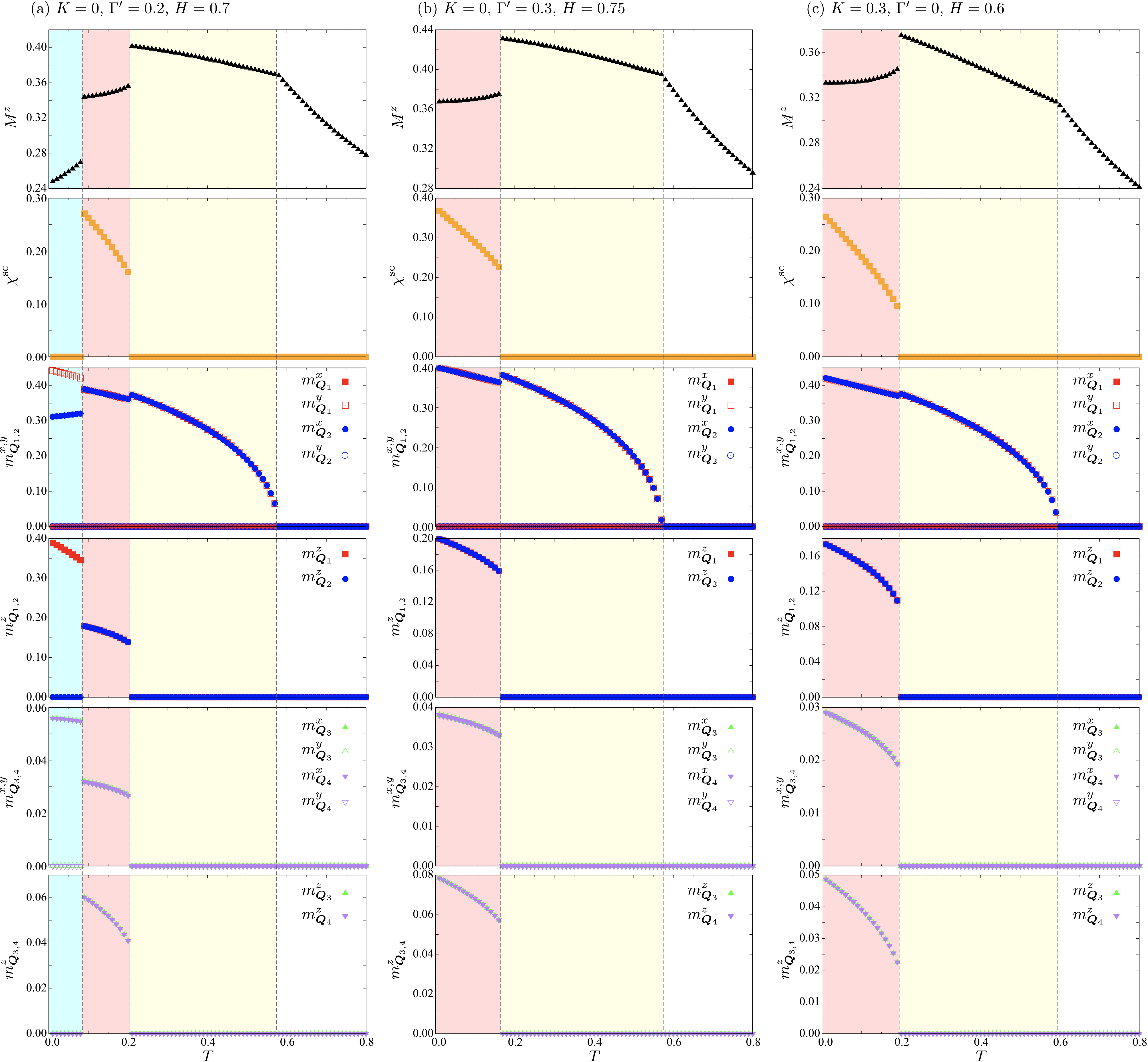} 
\caption{
\label{fig: Mq_Hz}
$T$ dependence of (first row) the magnetization $M^z$, (second row) the scalar chirality $\chi^{\rm sc}$, (third row) the in-plane magnetic moments at $\bm{Q}_{1,2}$, $m^{x,y}_{\bm{Q}_{1,2}}$, (fourth row) the out-of-plane magnetic moments at $\bm{Q}_{1,2}$, $m^{z}_{\bm{Q}_{1,2}}$, (fifth row) the in-plane magnetic moments at $\bm{Q}_{3,4}$, $m^{x,y}_{\bm{Q}_{3,4}}$, and (sixth row) the out-of-plane magnetic moments at $\bm{Q}_{3,4}$, $m^{z}_{\bm{Q}_{3,4}}$ at (a) $K=0$, $\Gamma'=0.2$, and $H=0.7$, (b) $K=0$, $\Gamma'=0.3$, and $H=0.75$, and (c) $K=0.3$, $\Gamma'=0$, and $H=0.6$. 
The vertical dashed lines represent the phase boundaries. 
}
\end{center}
\end{figure*}

\begin{figure*}[t!]
\begin{center}
\includegraphics[width=1.0 \hsize ]{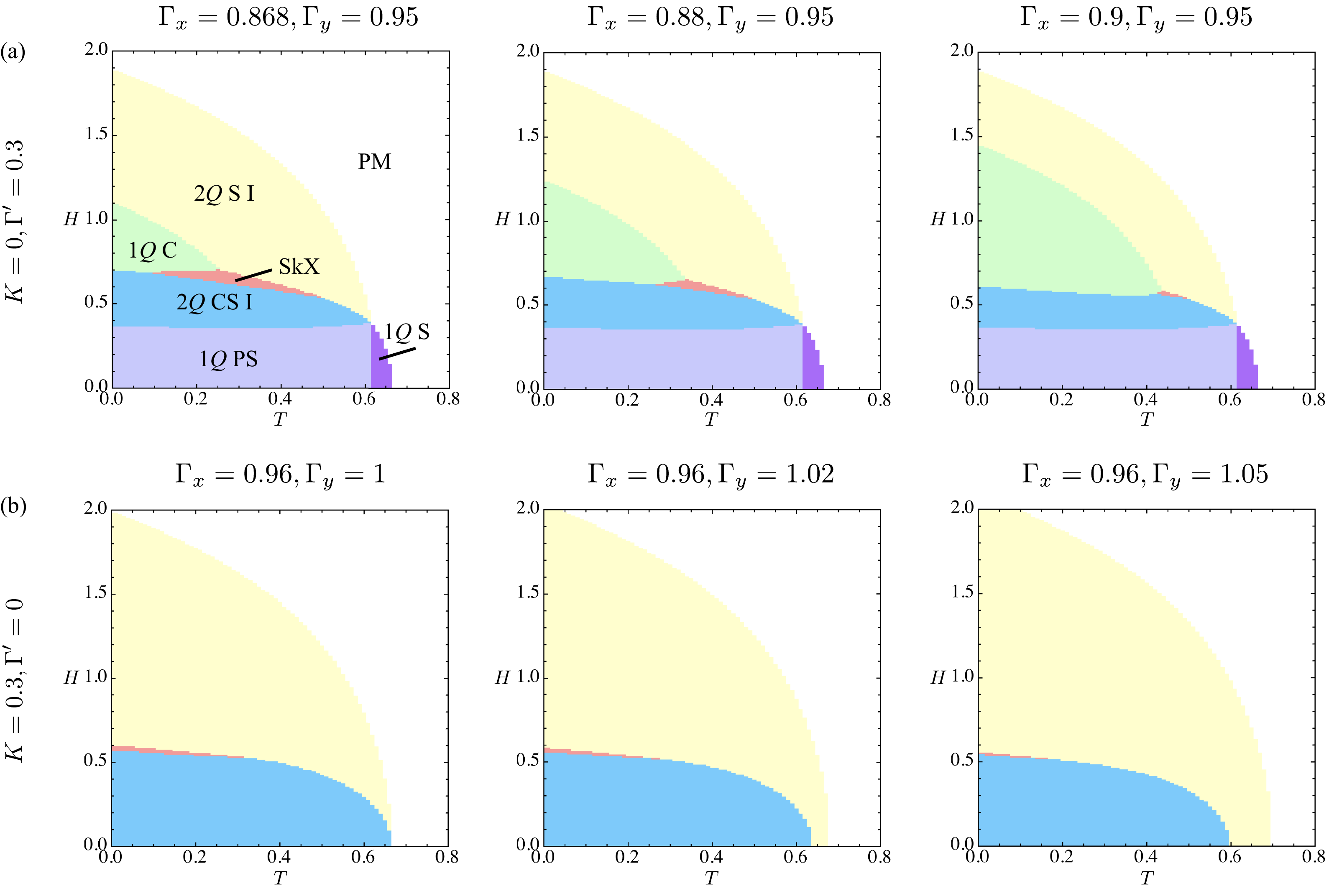} 
\caption{
\label{fig: PD_Hz_diff}
$H$--$T$ phase diagram for several values of $\Gamma_x$ and $\Gamma_y$ at (a) $K=0$ and $\Gamma'=0.3$, and (b) $K=0.3$ and $\Gamma'=0$. 
}
\end{center}
\end{figure*}

We present the $T$ dependence of the magnetization $M^z$, the scalar chirality $\chi^{\rm sc}$, and the $x$, $y$, and $z$ components of magnetic moments at $\bm{Q}_1$--$\bm{Q}_4$, $m^\alpha_{\bm{Q}_\nu}$, at $K=0$, $\Gamma'=0.2$, and $H=0.7$ in Fig.~\ref{fig: Mq_Hz}(a), and $K=0$, $\Gamma'=0.3$, and $H=0.75$ in Fig.~\ref{fig: Mq_Hz}(b). 
The SkX only appears at finite temperatures in Fig.~\ref{fig: Mq_Hz}(a), while it is stabilized from zero to finite temperatures in Fig.~\ref{fig: Mq_Hz}(b). 
In both figures, there is a clear jump in each quantity between the SkX and $2Q$ CS I (or 2$Q$ S I), which clearly indicates the first-order phase transition. 

The emergence of the SkX is attributed to the interplay between $\Gamma'$, the easy-axis anisotropy $\Gamma_z > \Gamma_{x,y}$, and the bond-dependent anisotropy $\Gamma_x \neq \Gamma_y$. 
To demonstrate that, we show the $H$--$T$ phase diagrams at fixed $K=0$ and $\Gamma'=0.3$ but different $\Gamma_x=0.868$, $0.88$, and $0.9$ in Fig.~\ref{fig: PD_Hz_diff}(a). 
With increasing $\Gamma_x$, the region in the 1$Q$ C phase is extended, while those in the SkX and 2$Q$ S I phases are shrunk. 
This is because the isotropic Heisenberg interaction tends to favor the 1$Q$ C state without the intensities at higher-harmonic wave vectors due to $m^x_{\bm{Q}_1}=m^y_{\bm{Q}_1}$ or $m^x_{\bm{Q}_2}=m^y_{\bm{Q}_2}$. 
Furthermore, one finds that the SkX in the ground state is rapidly replaced by the 1$Q$ C state with increasing $\Gamma_x$, which also indicates that the instability toward the SkX is larger at finite temperatures than at zero temperature.  

From the energetic viewpoint, the stabilization of the SkX by $\Gamma'$ is reasonable. 
This is understood from the spin configuration of the SkX, which is approximately given by 
\begin{align}
\label{eq:2QSkX}
\bm{S}_j \propto 
\left(
    \begin{array}{c}
    \cos \bm{Q}_1 \cdot \bm{r}_{j}+  \cos \bm{Q}_2 \cdot \bm{r}_{j} \\
    \cos \bm{Q}_1 \cdot \bm{r}_{j} -  \cos \bm{Q}_2 \cdot \bm{r}_{j} \\
    a_z (\sin \bm{Q}_1\cdot \bm{r}_{j} +\sin \bm{Q}_2\cdot \bm{r}_{j})+ \tilde{M}_z
          \end{array}
  \right)^{\rm T}, 
\end{align}
where $a_z$ and $\tilde{M}_z$ are the variational parameters depending on the model parameters, such as the magnetic anisotropy and the magnetic field.
Owing to the normalization constraint in terms of the spin length, i.e., $|\bm{S}_i| = 1$, there are intensities not only at $\bm{Q}_1$ and $\bm{Q}_2$ but also at high-harmonic wave vectors at $\bm{Q}_3$ and $\bm{Q}_4$. 
This means that the interactions in the $\bm{Q}_3$ and $\bm{Q}_4$ channels tend to favor the SkX. 
In addition, it is noteworthy that the contribution from such high-harmonic wave vectors is coupled to that from $\bm{Q}_1$ and $\bm{Q}_2$ like $(\bm{S}_{\bm{0}}\cdot \bm{S}_{-\bm{Q}_3})(\bm{S}_{\bm{Q}_1}\cdot \bm{S}_{\bm{Q}_2})$ and $(\bm{S}_{\bm{0}}\cdot \bm{S}_{-\bm{Q}_4})(\bm{S}_{-\bm{Q}_1}\cdot \bm{S}_{\bm{Q}_2})$ in the presence of the magnetic field in the free energy due to $\bm{Q}_1+\bm{Q}_{2}-\bm{Q}_3=0$ and $-\bm{Q}_1+\bm{Q}_2-\bm{Q}_4=0$.  
From the orientations of ${\bm m}_{{\bm Q}_\nu}$ in Figs.~\ref{fig: Mq_Hz}(a) and \ref{fig: Mq_Hz}(b), we conclude that the couplings in the form of $S^z_{\bm{0}} S^z_{-\bm{Q}_3} (\bm{S}_{\bm{Q}_1}\cdot \bm{S}_{\bm{Q}_2})+{\rm c.c.}$ and $S^z_{\bm{0}} S^z_{-\bm{Q}_4} (\bm{S}_{-\bm{Q}_1}\cdot \bm{S}_{\bm{Q}_2})+{\rm c.c.}$, for example, are important for the stabilization. 
Despite the importance of the high-harmonic channels, we refer to the SkX as the double-$Q$ state rather than the four-$Q$ state, since the this phase is continuously connected by that for nonzero $K$ without $\Gamma'$ [Fig.~\ref{fig: PD_Hz}], as discussed in the subsequent section.

\subsection{Case of biquadratic interaction}
\label{sec: Case of biquadratic interaction_Hz}

For $\Gamma'=0$, the $H$--$T$ phase diagrams for different $K=0.1$--$0.4$ are shown in the leftmost panel of Fig.~\ref{fig: PD_Hz}. 
For nonzero $K$, the SkX appears in the intermediate-field region similar to nonzero $\Gamma'$, while two single-$Q$ states (1$Q$ C and 1$Q$ PS) tend to be destabilized for $K>0$; the 1$Q$ C state vanishes for $K \gtrsim 0.1$ and the 1$Q$ PS state vanishes for $K \gtrsim 0.3$. 
Especially, the zero-field phase at low temperatures becomes the 2$Q$ CS I state instead of the 1$Q$ PS state for $K>0$, which indicates that the double-$Q$ instability at zero field means the importance of the biquadratic interaction $K$. 
In addition, a double-$Q$ state denoted as 2$Q$ S II appears only at finite $T$ for $K \gtrsim 0.3$, whose region is sandwiched by the 2$Q$ CS I and $1Q$ S states. 
The spin configuration of the 2$Q$ S II state is described by a linear combination of the sinusoidal waves with $\bm{Q}_1$ and $\bm{Q}_2$, whose spin components are given by $y$ ($z$) and $z$ ($x$) components, respectively. 

The obtained SkX for $K>0$ exhibits similar spin and scalar chirality textures to that for $\Gamma'>0$. 
For example the $T$ dependence of the magnetic moments and the spin scalar chirality at $K=0.3$, $\Gamma'=0$, and $H=0.6$ in Fig.~\ref{fig: Mq_Hz}(c) are similar to those in Fig.~\ref{fig: Mq_Hz}(b). 
Nevertheless, there are two different points in their $H$--$T$ phase diagrams. 
One is that the instability toward the SkX occurs at finite temperatures for nonzero $\Gamma'$, while such a clear feature is not found for nonzero $K$. 
Except for the $H$--$T$ phase diagram for $K=0.1$ and $\Gamma'=0$, where the SkX only appears in the narrow field region, the region of the SkX becomes smaller with increasing $T$. 
Such a tendency is also found when changing the anisotropic exchange interactions $\Gamma_x$ and $\Gamma_y$; the high-temperature region of the SkX becomes narrower for larger $\Gamma_x$ and $\Gamma_y$ while keeping the ground-state SkX, as shown in Fig.~\ref{fig: PD_Hz_diff}(b). 
The other is the enhancement of the SkX phase when increasing $K$ and $\Gamma'$. 
Compared to $\Gamma'$, the $K$ dependence of the SkX region against $H$ is small. 
Such a difference is accounted for by the different origins of the SkX. 
In the case of $K$, all the double-$Q$ states at low temperatures, 2$Q$ CS I, SkX, and 2$Q$ S I, have an energy gain by $K$, since $K$ brings about the energy loss to form the single-$Q$ spin configuration~\cite{Hayami_PhysRevB.95.224424}.  
Meanwhile, in the case of $\Gamma'$, only the 2$Q$ CS I and the SkX show an energy gain by $\Gamma'$ by reflecting their nonzero amplitudes of $\bm{m}_{\bm{Q}_{3,4}}$. 
In other words, there is no energy gain by $\Gamma'$ in the 2$Q$ S I state. 
In fact, one finds that the SkX region is extended to the high-field region with increasing $\Gamma'$. 
One also notices that the amplitudes of $\bm{m}_{\bm{Q}_{3,4}}$ tend to be larger for nonzero $\Gamma'$ in Fig.~\ref{fig: Mq_Hz}(b) than those for nonzero $K$ in Fig.~\ref{fig: Mq_Hz}(c).
This suggests that the effective coupling like $S^z_{\bm{0}} S^z_{-\bm{Q}_3} (\bm{S}_{\bm{Q}_1}\cdot \bm{S}_{\bm{Q}_2})+{\rm c.c.}$ and $S^z_{\bm{0}} S^z_{-\bm{Q}_4} (\bm{S}_{-\bm{Q}_1}\cdot \bm{S}_{\bm{Q}_2})+{\rm c.c.}$ plays an important role in stabilizing the SkX in the presence of $K$ as same as the case of $\Gamma'$
described in Sec.~\ref{sec: Case of high-harmonic wave-vector interaction_Hz}.

\subsection{Case of both interactions}
\label{sec: Case of both interactions_Hz}

When taking into account both interactions, the SkX tends to be more stable compared to the individual case, as shown in Fig.~\ref{fig: PD_Hz}. 
Thus, both interactions play a role in stabilizing the SkX in an additive way. 
The overall behavior in each phase is common to that in Secs.~\ref{sec: Case of high-harmonic wave-vector interaction_Hz} and \ref{sec: Case of biquadratic interaction_Hz}. 
With increasing $\Gamma'$, the phase boundary between the SkX and 2$Q$ S I phases moves to the high-field region so as to make the SkX phase more robust, while there is almost no $\Gamma'$ dependence in the other phase boundaries. 
On the other hand, the single-$Q$ states are replaced by the double-$Q$ states with increasing $K$ due to energy gain discussed in the previous section. 
In addition, the SkX region is slightly extended for larger $K$.

\section{Phase diagram under in-plane field}
\label{sec: Phase diagram under in-plane field}

\begin{figure*}[t!]
\begin{center}
\includegraphics[width=1.0 \hsize ]{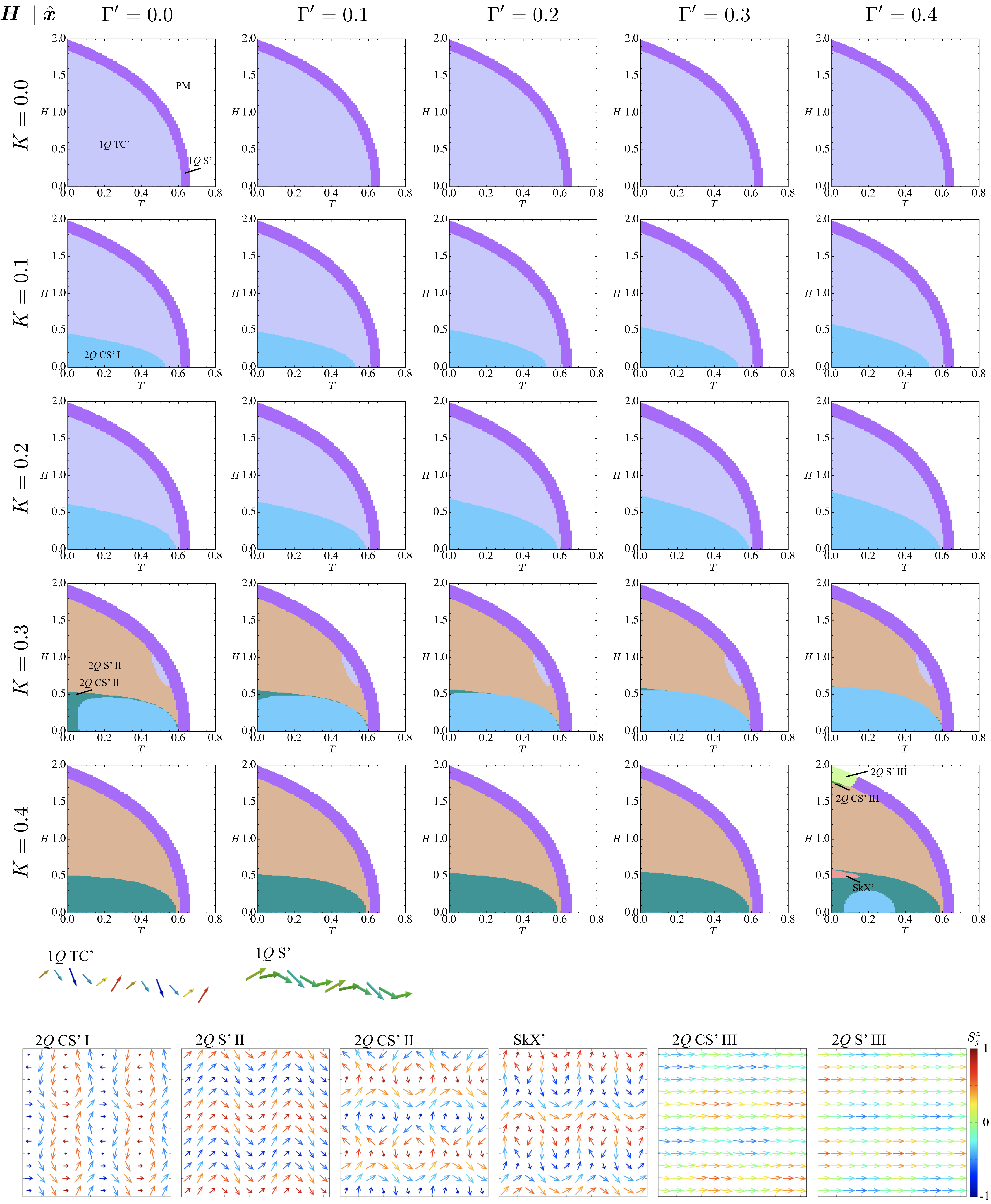} 
\caption{
\label{fig: PD_Hx}
$H$--$T$ phase diagrams of the model in Eq.~(\ref{eq: Ham}) for $\bm{H} \parallel \hat{\bm{x}}$ corresponding to Fig.~\ref{fig: PD_Hz}. 
TC represents transverse conical. 
}
\end{center}
\end{figure*}

\begin{table}[htb!]
\centering
\caption{
Scalar chirality $\chi^{\rm sc}$ and momentum-resolved magnetic moments $\bm{m}_{\bm{Q}_{\nu}}$ for $\nu=$1--4, in each magnetic phase for $\bm{H} \parallel \hat{\bm{x}}$. 
The subscript for $Q$ represents the indices for the ordering vector. 
The prime symbol for $Q$ represents the different magnitudes at $\bm{Q}_1$ and $\bm{Q}_2$. 
\label{table: OP_Hx}}
\renewcommand{\arraystretch}{2}
\begin{tabular}{lccccccccccc}\hline \hline
Phase  &$\chi^{\rm sc}$ & $m^{x}_{\bm{Q}_{1,2}}$ & $m^{y}_{\bm{Q}_{1,2}}$ & $m^{z}_{\bm{Q}_{1,2}}$ & $m^{x}_{\bm{Q}_{3,4}}$  & $m^{y}_{\bm{Q}_{3,4}}$ 
 & $m^{z}_{\bm{Q}_{3,4}}$ \\ \hline
1$Q$ TC' & -- & -- & $1Q_1$ & $1Q_1$ & --& -- & -- \\ 
1$Q$ S' & -- & -- & -- & $1Q$ & -- & -- & --\\ 
2$Q$ CS' I & -- & $1Q_2$ & $1Q_1$ & $1Q_1$ & -- & $2Q$ & $2Q$\\ 
2$Q$ CS' II & -- & $1Q_2$  & $1Q_1$ & $1Q_2$ & -- & $2Q$ & --\\ 
2$Q$ CS' III & -- & --  & $1Q_1$ & $2Q'$ & $2Q$ & -- & --\\ 
2$Q$ S' II & -- & -- & $1Q_1$ & $1Q_2$ & -- & -- & --\\ 
2$Q$ S' III & -- & -- & -- & $2Q$ & $2Q$ & -- & --\\ 
SkX' & $\checkmark$ & $2Q'$ & $1Q_1$ & $1Q_2$ & $2Q$ & $2Q$ & $2Q$\\ 
\hline \hline
\end{tabular}
\end{table}

We present the $H$--$T$ phase diagrams when applying the magnetic field along the $x$ direction, i.e., $\bm{H}=(H,0,0)$. 
We show the same plots as Fig.~\ref{fig: PD_Hz} under the in-plane field in Fig.~\ref{fig: PD_Hx}.  
In the $H$--$T$ phase diagrams produced for $\Delta K=0.1$ and $\Delta \Gamma'=0.1$, eight magnetic phases emerge with decreasing the temperature from the paramagnetic state at high temperatures. 
The nonzero components of the magnetic moments at $\bm{Q}_1$--$\bm{Q}_4$ in each phase are summarized in Table~\ref{table: OP_Hx}. 
Besides, the real-space spin configuration in each phase is shown in the bottom panel of Fig.~\ref{fig: PD_Hx}. 
In contrast to the out-of-plane field in Sec.~\ref{sec: Phase diagram under out-of-plane field}, the instability toward the SkX denoted as SkX' occurs only at $K=0.4$ and $\Gamma'=0.4$. 
Here and hereafter, the prime symbol means the magnetic phase under the in-plane field. 
Thus, larger biquadratic and high-harmonic wave-vector interactions are required to stabilize the topological spin textures under the in-plane field. 

At $\Gamma'=K=0$, there are only two magnetic phases in the phase diagram: One is the single-$Q$ transverse conical (1$Q$ TC') state and the other is the single-$Q$ sinusoidal (1$Q$ S') state. 
The 1$Q$ TC' state is characterized by the spiral wave along the $\bm{Q}_1$ direction, whose spiral plane lies on the $yz$ plane. 
With increasing $H$ and $T$, the $y$-spin modulation becomes zero while remaining the $z$-spin modulation, which means the appearance of the 1$Q$ S' state. 
In the end, there is no double-$Q$ instability under the in-plane field, which is different from the out-of-plane field. 
Qualitatively the same phase diagrams are obtained even for finite $\Gamma'$ at least up to $\Gamma'=0.4$ as shown in the top panel of Fig.~\ref{fig: PD_Hx}. 

When considering $K$ but $\Gamma'=0$, the double-$Q$ chiral stripe (2$Q$ CS' I) state appears in the low-field region for $K \gtrsim 0.1$ as shown in the leftmost panel of Fig.~\ref{fig: PD_Hx}. 
Similar to the 2$Q$ CS I under the out-of-plane field, the 2$Q$ CS' I state has double-$Q$ modulations consisting of the spiral wave along the $\bm{Q}_1$ direction and the sinusoidal wave along the $\bm{Q}_2$ direction. 
With further increasing $K$, the 1$Q$ TC' phase in the intermediate-to-high field regions is replaced by the double-$Q$ sinusoidal II (2$Q$ S' II) state for $K \gtrsim 0.3$.  
The spin configuration in this state consists of the two sinusoidal waves with $m^y_{\bm{Q}_1}$ and $m^z_{\bm{Q}_2}$. 
In addition, in the low-field region at low temperatures, the 2$Q$ CS' I state is replaced by the 2$Q$ CS' II state, where the sinusoidal modulation changes from $\bm{Q}_2$ to $\bm{Q}_1$. 
At $K=0.4$, the 2$Q$ CS' I is completely replaced by the 2$Q$ CS' II state.

For both nonzero $\Gamma'$ and $K$, the overall $H$--$T$ phase diagrams are qualitatively similar with changing $\Gamma'$ for fixed $K$ when $K$ is small.
At $K=0.3$, the 2$Q$ CS' II state is replaced by the 2$Q$ CS' I state when $\Gamma'$ is increased.
At $\Gamma'=0.4$ and $K=0.4$, in addition to the SkX', two double-$Q$ states denoted as 2$Q$ CS' III and 2$Q$ S' III appear in the high-field region, which are characterized by different double-$Q$ superpositions as summarized in Table~\ref{table: OP_Hx}. 
Among all the obtained phases, only the SkX' exhibits a nonzero scalar chirality.

\section{Comparison with skyrmion-hosting materials}
\label{sec: Comparison with skyrmion-hosting materials}

Finally, let us compare the $H$--$T$ phase diagrams of the effective spin model in the wide range of the model parameters with that observed in the SkX-hosting tetragonal material GdRu$_2$Si$_2$~\cite{khanh2020nanometric, Yasui2020imaging,khanh2022zoology}. 
In experiments, there are three magnetic phases denoted as Phase I, Phase II, and Phase III in the out-of-plane field direction, while there are four magnetic phases denoted as Phase I, Phase IV, Phase III', and Phase V in the in-plane field direction~\cite{khanh2022zoology}. 
Based on the resonant x-ray scattering and spectroscopic-imaging scanning tunneling microscopy measurements, each phase was identified as follows~\cite{khanh2020nanometric, Yasui2020imaging,khanh2022zoology}: 2$Q$ CS I (and 2$Q$ CS' I) for Phase I, the SkX for Phase II, 2$Q$ S I for Phase III, 1$Q$ TC' for Phase IV, 2$Q$ S' II for Phase III', and 1$Q$ S' for Phase V. 

First, let us consider the case under the out-of-plane field. 
As the zero-field state at low temperatures corresponds to the 2$Q$ CS I state, the biquadratic interaction should be nonzero in this compound. 
Indeed, for nonzero $K$, the emergence of three phases with changing $H$ at low temperatures is consistent with the experimental observations. 
Moreover, the fragility of the SkX against thermal fluctuations compared to the 2$Q$ CS I and 2$Q$ S I states is well reproduced in the effective spin model. 
It is noted that a large value of $K$ is not necessary to realize such a phase sequence by taking into account $\Gamma'$. 
For example, the stability region of the SkX phase at $K=0.3$ and $\Gamma'=0$ is similar to that at $K=0.1$ and $\Gamma'=0.1$. 

In addition, by focusing on the high-temperature region for nonzero $K$, additional phases denoted as the 1$Q$ S and 2$Q$ S II appear depending on $K$ in the effective spin model, as discussed above. 
Notably, the magnetization measurement and the resonant x-ray scattering measurement implied the emergence of the 2$Q$ S II phase~\cite{khanh2020nanometric, khanh2022zoology}. 
Although the appearance of the 1$Q$ S phase has not been clarified, our results based on the effective spin model indicate that the 1$Q$ S phase might additionally appear in the higher-temperature region next to the 2$Q$ S II phase. 

Next, let us compare the case under the in-plane field. 
As shown in the phase diagrams in Fig.~\ref{fig: PD_Hx}, we obtain the instabilities toward the magnetic states observed in GdRu$_2$Si$_2$, i.e., 2$Q$ CS' I, 1$Q$ TC', 2$Q$ S' II, and 1$Q$ S' with changing $K$ and $\Gamma'$. 
Thus, the effective spin model roughly reproduces the $H$--$T$ phase diagram in GdRu$_2$Si$_2$. 
Meanwhile, we could not obtain the phase sequence from the 1$Q$ TC' to the 2$Q$ S' II at low temperatures in the present model-parameter range, which was observed in experiments~\cite{khanh2022zoology}. 
Thus, further additional interactions and anisotropy might be required to realize such a phase sequence under the in-plane field.

\section{Summary}
\label{sec: Summary}

To summarize, we have investigated the magnetic field--temperature phase diagram of the effective spin model in itinerant centrosymmetric tetragonal magnets. 
By focusing on the two mechanisms to induce the SkX, the higher-harmonic wave-vector interaction and the biquadratic interaction, we construct the phase diagrams in a wide range of model parameters based on the efficient steepest descent method. 
As a result, we show the stability tendency of the SkX against these interactions as well as the magnetic field and temperature. 
Especially, we find that the instability toward the SkX under the out-of-plane magnetic field occurs at finite temperatures by the higher-harmonic wave-vector interaction, while that occurs in the ground state by the biquadratic interaction. 
Furthermore, we reveal the tendency of the other single-$Q$ and double-$Q$ phases in both in-plane and out-of-plane magnetic fields, which provides information about the microscopic important interactions. 
We also discuss the relevance of our results to the experimental phase diagram in GdRu$_2$Si$_2$. 
Based on the obtained phase diagram in the effective spin model, we conclude that the biquadratic interaction plays an important role and propose the additional phase at high temperatures. 
Our systematic investigation of the magnetic field--temperature phase diagrams would be a useful reference to construct the effective spin model for the materials hosting the multiple-$Q$ states in the centrosymmetric tetragonal magnets, such as EuAl$_4$~\cite{Shang_PhysRevB.103.L020405,kaneko2021charge,takagi2022square,Zhu_PhysRevB.105.014423}. EuGa$_4$~\cite{zhang2022giant, Zhu_PhysRevB.105.014423}, EuGa$_2$Al$_2$~\cite{moya2021incommensurate}, Mn$_{2-x}$Zn$_x$Sb~\cite{Nabi_PhysRevB.104.174419}, and MnPtGa~\cite{ibarra2022noncollinear}.

\begin{acknowledgments}
S.H. thank S. Seki for fruitful discussions. 
S.H. acknowledges Y. Motome for enlightening discussions in the early stage of this study.
This research was supported by JSPS KAKENHI Grants Numbers JP21H01037, JP22H04468, JP22H00101, JP22H01183, JP22K03509, and by JST PRESTO (JPMJPR20L8). 
Parts of the numerical calculations were performed in the supercomputing systems in ISSP, the University of Tokyo.
\end{acknowledgments}

\bibliographystyle{apsrev}
\bibliography{ref}

\end{document}